\documentclass[%
 reprint,
 amsmath,amssymb,
 aps,
 superscriptaddress
]{revtex4-2}

\usepackage{graphicx}
\usepackage{dcolumn}
\usepackage{bm}
\usepackage{hyperref,color}
\usepackage{sidecap}

\begin{document}

\preprint{APS/123-QED}

\title{Bloch-wave Interferometry of
Driven Quasiparticles in Bulk GaAs}

\author{Seamus D. O'Hara}
\email{ohara@physics.ucsb.edu}
\affiliation{Physics Department, University of California, Santa Barbara, California 93106, USA}
\affiliation{Institute for Terahertz Science and Technology, University of California, Santa Barbara, California 93106, USA}
\author{Joseph B. Costello}
\affiliation{Physics Department, University of California, Santa Barbara, California 93106, USA}
\affiliation{Institute for Terahertz Science and Technology, University of California, Santa Barbara, California 93106, USA}
\author{Qile Wu}
\affiliation{Physics Department, University of California, Santa Barbara, California 93106, USA}
\affiliation{Institute for Terahertz Science and Technology, University of California, Santa Barbara, California 93106, USA}

\author{Ken West}
\affiliation{Electrical Engineering Department, Princeton University, Princeton, New Jersey 08544, USA}

\author{Loren Pfeiffer}
\affiliation{Electrical Engineering Department, Princeton University, Princeton, New Jersey 08544, USA}

\author{Mark S. Sherwin}
\email{sherwin@physics.ucsb.edu}
\affiliation{Physics Department, University of California, Santa Barbara, California 93106, USA}
\affiliation{Institute for Terahertz Science and Technology, University of California, Santa Barbara, California 93106, USA}

\date{\today}

\begin{abstract}
We report that the polarizations of sidebands emitted from bulk gallium arsenide (GaAs) driven by a strong terahertz (THz) laser while probed with a weak near-infrared laser can be viewed as interferograms from a Michelson-like interferometer for Bloch waves. A simple analytical model is introduced to calculate the difference in quantum mechanical phases accumulated by Bloch waves associated with electron-heavy hole and electron-light hole pairs in their respective interferometer arms. The measured and calculated spectra are in good quantitative agreement, including scaling with THz field strength. Our results indicate a simple way to extract material parameters in future experiments.

\end{abstract}

\maketitle

Since Thomas Young demonstrated the wave nature of light through interference phenomena, physicists have exploited interferometry in many forms to measure phase differences between waves.  For example, A. A. Michelson invented the interferometer that bears his name to reject the hypothesis of a stationary luminiferous ether ~\cite{Michelson}. After the advent of quantum mechanics, Thomson and Davisson demonstrated the wave nature of free electrons~\cite{Thomson,DavissonAndGermer}. Decades later, a Young's-type double-slit experiment demonstrated the wave nature of cold atoms  ~\cite{1991YoungInterferometer}. The dynamical phase of a matter wave is sensitive to its acceleration, which has been exploited to measure inertial forces by interfering atoms evolving over two quantum paths ~\cite{AtomGravity,FreqComInt}. In the solid state, standing waves formed by the interference of Bloch-electron waves on metal surfaces have been observed ~\cite{1993StandingWaves,1993IBM,1994scattering}, and transport phenomena like weak localization are manifestations of interference in the presence of disorder~\cite{weaklocalization}. In these near-equilibrium systems the electrons do not accelerate. In driven solids, rapid scattering processes complicate the measurement of electrons' time-dependent quantum mechanical phases, requiring a combination of strong fields and ultrafast probes to measure the effects of electron acceleration.

The recent development of strong laser fields has allowed coherent acceleration of electrons on picosecond and sub-picosecond time scales~\cite{2001HHG,XUV}. In high-harmonic generation (HHG) in solids, a single strong laser field creates and then accelerates charged excitations ~\cite{HHGgasesAndSolids,HHGsolids,HHGsolidReview,HHGinSCmaterials,HHGsolidTheory,RupertHHG}. HHG enables studies of electron dynamics with attosecond resolution, ~\cite{UltrafastMetrology,AttoSShhg}, and is sensitive to many details of the interactions of quasiparticles with their host crystals ~\cite{AnisoHHGbulkMgO,AttoSShhg,HHGMoS2,HHGellip,UltrafastMetrology,LDrivenBandStHHG}. In 2015, time-resolved interferences of electronic waves were observed in HHG in a semiconductor that was driven by strong terahertz pulses ~\cite{RupertHHG}. However, such quantum interferences in HHG cannot be easily attributed to a small number of momentum-space trajectories for the Bloch waves. In an interferometry picture, where, for example, each excited Bloch wave exists in a different arm of a Michelson interferometer, the HHG process would be mapped onto an interferometer with many, many arms ~\cite{prlHHGInt}, making it difficult to connect a given signal to a particular characteristic of the many excited Bloch waves.

\begin{figure}
\includegraphics[scale = 0.97]{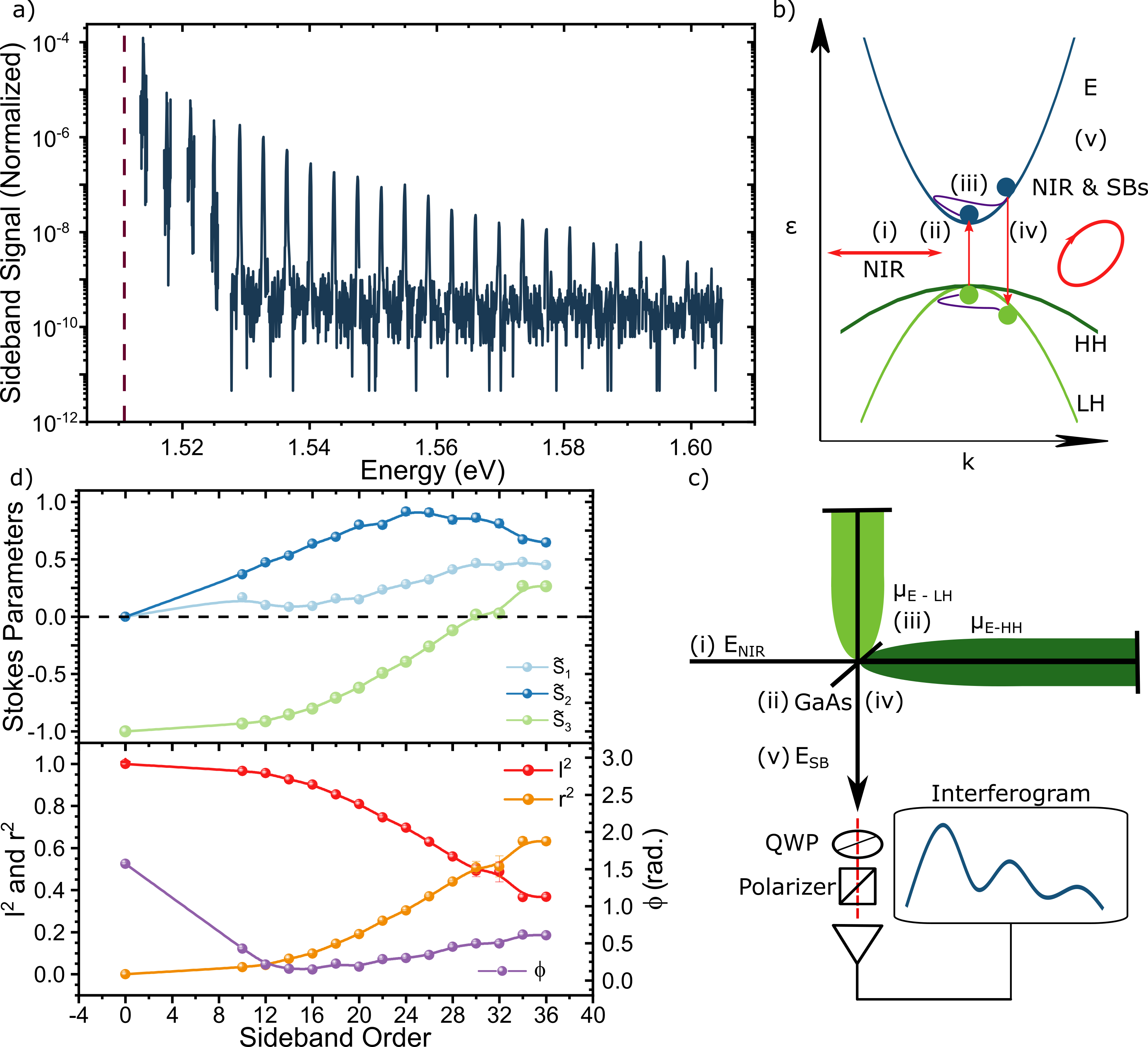}
\caption{\label{fig:Concept} 
Description of high-order sideband generation (HSG) in bulk GaAs as a Michelson interferometer for Bloch waves. (a) An example of a measured sideband spectrum with NIR laser frequency at red dashed line. (b) A representation of HSG in momentum space showing electron (E), heavy-hole (HH), and light-hole (LH) bands. (c) A Michelson interferometer for Bloch waves. The Roman numerals in (b) and (c) depict the following processes. (i) A NIR laser is incident on the bulk GaAs. (ii) The bulk GaAs acts like a “beam-splitter”, converting the NIR laser beam into E-HH and E-LH pairs (iii). In the two arms of the interferometer, the Bloch waves of the E-HH and E-LH pairs propagate along different k-space trajectories and acquire different phases. (iv) These Bloch waves merge at the beam-splitter and (v) sidebands are emitted. Bloch wave interferograms of sideband polarizations are recorded by Stokes polarimetry using a quarter-wave plate (QWP) and a polarizer. (d) Stokes polarimetry and a Bloch wave interferogram. Top frame: Three normalized Stokes parameters $\Tilde{S}_1$, $\Tilde{S}_2$ and $\Tilde{S}_3$ corresponding to the sideband spectra in (a). Bottom frame: Bloch wave interferograms, consisting of the sideband polarization as a function of sideband order $n$. The polarization state of the $n$th-order sideband is represented as a normalized state in the basis of circular polarizations, $|E^{\rm SB}_{n} \rangle= l(n)e^{i\varphi(n)}|L\rangle+ r(n)|R\rangle$, where $l(n)$, $r(n)$, and $\varphi(n)$ are real functions of $n$ \cite{SI}}
\end{figure}

In this Letter, we demonstrate that a related process, high-order sideband generation (HSG),  \cite{FirstHSG,HSGBulkGaAs,QuantWellHSG,Rupert1HSG,HSGtmdc,DarrenHSG,HSGattoBlochE} can be viewed as the output of a Michelson interferometer for Bloch waves associated with electron-hole pairs. In HSG two lasers with different frequencies are used. A relatively weak laser tuned near the band gap of a semiconductor --- bulk Gallium Arsenide (GaAs) for the experiments presented here --- creates electron-hole pairs, and a second, stronger, lower-frequency laser accelerates them to higher energy. The use of two frequencies separates the contributions of intraband excitations and interband dynamics in the final HSG signal \cite{Qile}. The Bloch-wave interferograms are sideband polarizations as functions of sideband photon energy.

An example HSG spectrum is displayed in Fig. ~\ref{fig:Concept} (a). Each peak represents a sideband with a photon energy $\hbar\Omega_{\rm NIR}+n\hbar\omega$, where $\hbar\Omega_{\rm NIR}$ and $\hbar\omega_{\rm THz}$ are respectively the photon energies of the near-infrared (NIR) and terahertz (THz) lasers, and the sideband order $n$ is an even integer. We tune the NIR laser close to the estimated bandgap~\cite{SI} with a wavelength of 820.6 nm. The THz field has a frequency $f_{\rm THz}=447 \pm 1$ GHz ~\cite{DarrenHSG} and a pulse duration of 40 ns. The experiment was performed at 35 K. Like HHG, HSG is summarized in three steps \cite{FirstHSG}, displayed in a  momentum space, semiclassical picture in Fig. ~\ref{fig:Concept} (b). First, a NIR laser creates an electron-hole pair, where the hole is in a superposition of heavy-hole (HH) and light-hole (LH) states. Second, a linearly-polarized THz field drives the electron-hole (E-H) pairs towards higher quasimomenta. Third, the electron and hole recombine to generate a sideband photon.

The microscopic processes of HSG described in Fig.\ref{fig:Concept}(b) have been labeled i to v to map them onto different components of a Michelson interferometer for Bloch waves, shown in Fig. \ref{fig:Concept}(c). The NIR laser is incident on the bulk GaAs (i), which acts like a beam-splitter (ii), creating E-H states in one or both legs. Under the THz field, the E-HH and E-LH Bloch waves propagate along different k-space trajectories (iii). Because the E-HH and E-LH reduced masses determine the dispersion relations for Bloch waves, they are analogous to refractive indices that determine the dispersion relations for light waves in the arms of a conventional Michelson interferometer. The effective arm lengths in the Michelson interferometer for Bloch waves, which determine the quantum mechanical phases acquired by E-HH and E-LH pairs, increase with increasing sideband order for a fixed THz field. Upon sideband emission, the E-HH and E-LH Bloch waves merge at the beam- splitter (iv), coupling out of the interferometer. Bloch-wave interferograms  \cite{BlochWave} of the outgoing sideband electric field (v) consist of the sideband polarizations as functions of sideband order. 

We record our Bloch-wave interferograms using Stokes polarimetry, where all four Stokes parameters for each sideband are measured~\cite{SI}. The top frame of Fig.~\ref{fig:Concept} (d) shows three Stokes parameters, $S_1$, $S_2$, and $S_3$ normalized as $\tilde{S}_j=S_j/\sqrt{S_1^2 + S_2^2 + S_3^2}$ ($j=1,2,3$). The corresponding Bloch-wave interferograms are shown in the bottom frame, where the data points for $n=0$ represent the polarization state of the NIR laser. The polarization state of the $n$th-order sideband is represented as a normalized state in the basis of circular polarizations, $|E_{\rm SB,n} \rangle= l(n)e^{i\phi(n)}|L\rangle+ r(n)|R\rangle$. The functions $l(n)$, $r(n)$, and $\phi(n)$ are real and satisfy
\begin{align}
        \phi(n) & =\tan^{-1}\left(\frac{S_2(n)}{S_1(n)}\right)-2\theta-\frac{\pi}{2}, \\
        l^2(n)&= \frac{1-\tilde{S}_3(n)}{2}=1-r^2(n),
        \label{Stokes2lrPhi}
\end{align}
where $\theta$ is the angle between the crystal axis [110] and the linear polarization of the THz field. The left-handed and right-handed circularly polarized states, $|L\rangle$ and $|R\rangle$, are defined with respect to the crystal axes of bulk GaAs~\cite{SI}.

Following Ref.~\cite{BlochWave}, the relation between the electric fields of the $n$th-order sideband and the NIR laser, ${\bf E}_{\rm SB,n}$ and ${\bf E}_{\rm NIR}$, can be written as
\begin{align}
    {\bf E}_{\rm SB,n}
    \propto
    	\sum_{s=1,2}
	&
        		\begin{pmatrix}
	         {\bf d}_{\rm E-HH,s}\\
        		 {\bf d}_{\rm E-LH,s}
	        \end{pmatrix}^{\dag}
	        		\begin{pmatrix}
	         		\varsigma^{\rm{HH}}_n  & 0\\
	         		0  &  \varsigma^{\rm{LH}}_n       		 	
	       		 \end{pmatrix} 	      
	  \notag\\          		
	  &
	        \begin{pmatrix}
	         {\bf d}_{\rm E-HH,s}\\
        		 {\bf d}_{\rm E-LH,s}
	        \end{pmatrix} 
	\cdot
	{\bf E}_{\rm NIR},
 	\label{EQ:ESB_from_ENIR}
\end{align}
where $s$ labels the two-fold degeneracy in the electron-hole states, ${\bf d}_{\rm E-HH,s}$ (${\bf d}_{\rm E-LH,s}$) is the dipole vector associated with the E-HH (E-LH) states labeled by $s$, and $\varsigma^{\rm{HH}}_n$ ($\varsigma^{\rm{LH}}_n$) is a propagator describing the recollisions of the E-HH (E-LH) pairs. The elements i-v of the Michelson interferometer for Bloch waves shown in Fig. ~\ref{fig:Concept}(c) can be directly identified with the five terms of Eq.(\ref{EQ:ESB_from_ENIR}), reading from right to left.

\begin{figure*}[t!]
\includegraphics[scale = 1]{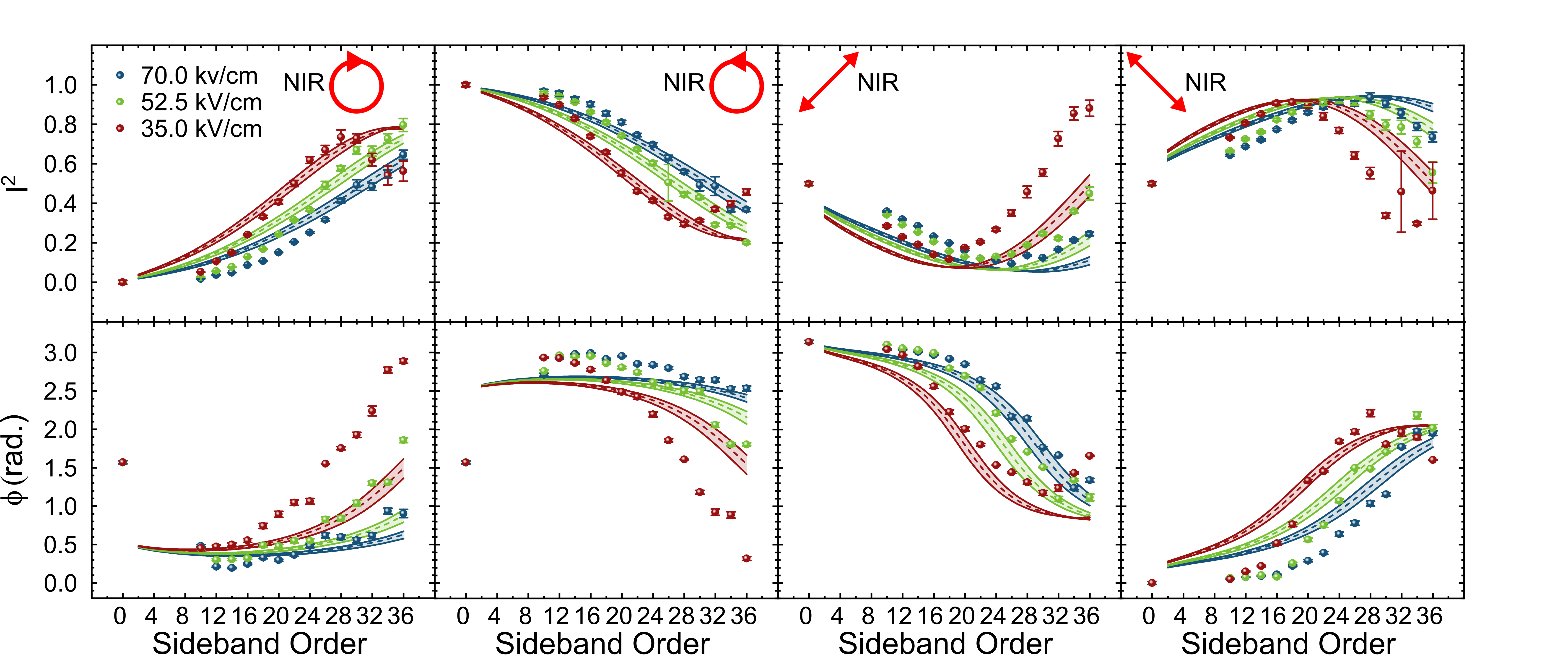}
\caption{\label{fig:MainData} 
Bloch wave interferograms for different NIR polarizations and THz field strengths. First row: the left-handed circularly polarized component ($l^2(n)$) (see Eq. \ref{Stokes2lrPhi}). Second row: the relative phase $\phi(n)$ between the left-handed and right-handed components of the sideband photons. Columns from left to right: four different polarization states of the NIR laser (cartoon in upper corner of top row). First (right-handed circularly polarized), second (left-handed circularly polarized), third (diagonal), and fourth (anti-diagonal). The data is taken with three THz field strength, 70.0 kV/cm (blue symbols), 52.5 kV/cm (dark green symbols), and 35.0 kV/cm (red symbols). The dashed lines represent the results calculated under the LIT approximation, with the bands representing a $\pm 7\%$ experimental error in field strength applied to the LIT calculations.}
\end{figure*}

In order to better understand the sideband polarizations, and inspired by a semiclassical description of the quasiparticle trajectories~\cite{SI,BlochWave}, we model the propagators $\varsigma^{\nu}_n$ $(\nu = \rm{HH}, \rm{LH})$ with the expression
\begin{align}
\varsigma_{n}^{\nu}= \exp[i(n\omega t_{f,n,\nu} + A_{n,\nu})+ (i\Gamma_{d} + \Delta_{\rm NIR})\frac{\tau_{n,\nu}}{\hbar})],
\label{varsigma}
\end{align}
where $\hbar$ is the reduced Planck constant, $\tau_{n,\nu} \equiv t_{f,n,\nu} - t_{o,n,\nu}$ is the time required for an E-H pair to be accelerated by the THz field to gain an energy offset $n\hbar\omega$, $\omega$ is the angular frequency of the THz field, $\Gamma_d$ is the dephasing constant of the E-H pairs, $\Delta_{\rm{NIR}} \equiv \hbar\Omega - E_g$ is the NIR detuning from the gap energy $E_g$, and $A_{n,\nu}$ is the dynamical phase of the E-H pair accumulated during the acceleration time $\tau_{n,\nu}$. Eq. \ref{varsigma} has an oscillating term $e^{i\Theta_{\nu}(n)}$, where $\Theta_{\nu}(n) = A_{n,\nu} + n \omega t_{f,n,\nu} + \Delta_{NIR}\tau_{n,\nu}\hbar^{-1}$. So these propagators oscillate as a function of sideband order with a phase that depends on $A_{n,\nu}$. The presence of $\exp(-\Gamma_d\tau_{n,\nu}\hbar^{-1})$ in Eq. \ref{varsigma} damps the oscillations of the propagators.

Even with this expression for $\varsigma_{n}^{\nu}$, for a sinusoidal THz field, $\mathbf{E}_{\rm{THz}}(t) = F_{\rm{THz}}\sin(\omega t)$, there is no analytic solution for $t_{o,n,\nu}$ and $t_{f,n,\nu}$. As a result there are no analytic expressions for the Bloch waves and their phases $\Theta_{\nu}(n)$, only numerical solutions. For experiments reported here, it is reasonable to approximate the THz field as linear in time (LIT)~\cite{Qile}, because all sidebands arise from E-H pairs accelerated within less than 300 fs of a zero-crossing of the THz field, which has a period of 2.2 ps ~\cite{SI}. With the THz field approximated as LIT, $E_{\rm{THz}} = F_{\rm{THz}}\omega t$, the following analytical expression for $t_{o,n,\nu}$ and $t_{f,n,\nu}$ can be derived from simple kinematics:
\begin{equation}\label{LinFieldTimes}
    -2\omega t_{o,n,\nu} = \omega t_{f,n,\nu} = \frac{2}{\sqrt{3}}\left(\frac{8n\hbar\mu_{\nu}\omega^3}{e^2F_{\rm{THz}}^2}\right)^{1/4}.
\end{equation}
Here n is the sideband order, $\mu_{\nu}$ is the reduced mass of the e-h pair in bulk GaAs \cite{LuttHam,Vurgaftman}, $e$ is the fundamental charge, and $F_{\rm{THz}}$ is the peak THz field strength.

In the LIT approximation, we can calculate values for the position, energy, and dynamical phase of a quasiparticle during its acceleration by the THz field~\cite{SI}. In this limit, the expression for $A_{n,\nu}$ becomes
\begin{equation}\label{DynamicalPhase}
    A_{n,\nu} = -\frac{2\sqrt{3}}{15}\left(\frac{8n^5\hbar\omega^3\mu_{\nu}}{e^2F_{THz}^2}\right)^{1/4},
\end{equation}
which illustrates the dynamical phase’s explicit dependence on different experimental variables, and allows us to calculate the dynamics of the Bloch waves in our system. Using Eq. \ref{EQ:ESB_from_ENIR}-\ref{DynamicalPhase}, we can predict the outgoing sideband polarization from an arbitrary NIR electric field. We note that, in the LIT approximation, both the terms $n\omega t_{f,n,\nu}$ and $A_{n,\nu}$ in Eq. \ref{varsigma} scale as $({n^5/{F_{\rm{THz}}^2}})^{1/4}$.

\begin{figure*}
\includegraphics[scale = 1]{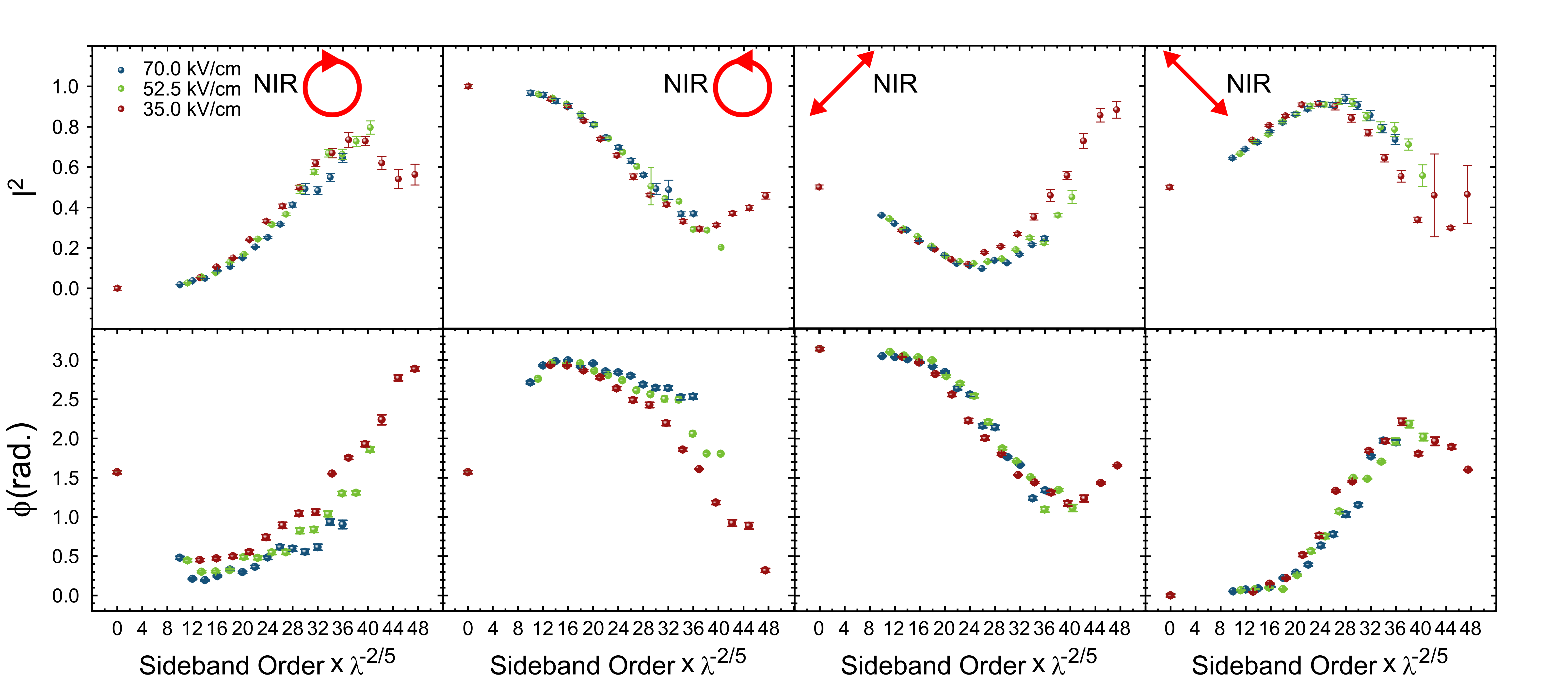}
\caption{\label{fig:ScaledData} 
Scaled Bloch wave interferograms. The experimental data shown in Fig.~\ref{fig:MainData} with $l^2(n)$ and $\phi(n)$ plotted as functions of a rescaled sideband order, $n\times\lambda^{-2/5}$ ($\lambda=F_{\rm THz}/(70$ kV/cm)), the scaling factor predicted by the LIT approximation and Eq. \ref{DynamicalPhase} to have the same dynamical phase at each x-value}
\end{figure*}

In Figure \ref{fig:MainData}, we display polarization data, or Bloch wave interferograms, taken with different THz field strengths and NIR polarization conditions. The top row is the measured $l(n)^2$ recorded at each sideband order. The bottom row is the measured $\phi(n)$. A different NIR polarization is used in each column, depicted with a cartoon in the upper corner. Data were taken at $F_{\rm{THz}}$ = 70.0, 52.5, and 35.0 kV/cm. Qualitatively, the data demonstrate the behavior predicted by Eq. \ref{varsigma}, with $l(n)^2 \text{ and } \phi(n)$ changing with sideband order like a damped sine wave. The lower field strength data curves oscillate more rapidly with sideband order, following the $F_{\rm{THz}}^{-1/2}$ dependence of $\tau_{n,\nu}$ and $A_{n,\nu}$ of Eqs. \ref{LinFieldTimes} and \ref{DynamicalPhase}. The dashed lines in each of the plots are the sideband polarizations predicted from Eq \ref{EQ:ESB_from_ENIR}, using the literature values of the Luttinger Hamiltonian \cite{Vurgaftman} in bulk GaAs:$\gamma_1= 6.85, \gamma_2 = 2.10, \gamma_3 = 2.90$. The NIR detuning $\Delta_{\rm{NIR}}$ was set to 0~\cite{SI}. Only the parameter $\Gamma_d$ was adjusted to calculate the dashed lines. The value $\Gamma_d = 4.8 \hbar\omega$ maximizes quantitative agreement between the calculated polarizations and experiment and was used to calculate all 24 dashed lines in Fig \ref{fig:MainData}. The positive and negative bands of error represent the $\pm 7\%$ error in THz field strength present in experiment.

Finally, to illustrate the roles of dynamical phase and recollision time in the Bloch wave interferograms, we rescale the data in Fig. \ref{fig:MainData} so that each point on the x-axis is proportional to $\Theta_{\nu}(n)$, independent of THz field strength. For $\Delta_{NIR}=0$, $\Theta_{\nu}(n)=A_{n,\nu}+n\omega t_{f,n,\nu}$. Examination of Eqs. \ref{LinFieldTimes} and \ref{DynamicalPhase} shows that both $A_{n,\nu}$ and $n\omega t_{f,n,\nu}$ scale as $({{n^5}/{{F_{\rm{THz}}}^2}})^{1/4}$. Thus, if the LIT approximation holds, and the NIR laser is tuned sufficiently close to the band gap that $\Delta_{NIR} \approx 0$, we expect that $\Theta_{\nu}(n)$ should also scale as $({{n^5}/{{F_{\rm{THz}}}^2}})^{1/4}$. To check this on data sets taken with different THz field strengths (see Fig. \ref{fig:MainData}), we multiply the sideband order $n$ for each sideband  by $\lambda ^{-2/5}$, where $\lambda \equiv F_{\rm{THz}}/(70 \text{kV}\cdot \text{cm}^{-1})$. Fig. \ref{fig:ScaledData} shows that the Bloch wave interferograms measured with different $F_{\rm{THz}}$ collapse onto a single curve, in agreement with the predicted LIT scaling law. 

The success of the scaling shown in Fig. \ref{fig:ScaledData}, together with the agreement between calculated and measured Bloch wave interferograms in Fig. ~\ref{fig:MainData}, strongly support our assertion that the HSG process in bulk GaAs can be viewed as a Michelson interferometer for Bloch waves. In this interferometer, both the species-dependent recollision times $t_{f,n,\nu}$  and dynamical phases $A_{n,\nu}$ contribute to the phase difference acquired between Bloch waves associated with E-LH and E-HH pairs during the acceleration phase. The E-LH pairs, because of their smaller reduced mass, take less time to reach the same kinetic energy as the heavier E-HH pairs (see Eq. \ref{LinFieldTimes}), and also acquire a smaller dynamical phase (see Eq. \ref{DynamicalPhase}). 

We note that, for the RHCP and LHCP excitation, the calculated value of $l^2(n)$ agrees better with the experimental data than $\phi(n)$ (see Fig. \ref{fig:MainData}), and the $l^2(n)$ curves collapse more completely upon scaling than the $\phi(n)$ curves (see Fig. \ref{fig:ScaledData}). The opposite is true for the calculated and scaled $l^2(n)$ and $\phi(n)$ for the diagonal and antidiagonal excitation. Part of this difference may be due to the initial conditions seeded by the NIR pulse. For the RHCP and LHCP excitation, we are effectively seeding the initial LHCP and RHCP components of the system, and the phase delay $\phi(\rm{NIR})$ between RHCP and LHCP of the NIR laser is not defined. In the diagonal and antidiagonal excitation, we are intentionally starting with a particular phase delay $\phi(\rm{NIR})$ between the LHCP and RHCP components of the NIR field. In all four of these instances, the variable we more intentionally manipulated with experiment has a better quantitative agreement between experiment and theory. By taking into account the effects of quantum fluctuations and going beyond the LIT approximation as in Ref.\cite{Qile}, and by allowing additional free parameters---for example, detuning $\Delta_{\rm{NIR}}$, $\mu_\nu$ different from those predicted by literature values of the Luttinger parameters, dephasing that varies with sideband order, and the influence of unavoidable small strains that lift the degeneracy between HH and LH at the center of the Brillouin zone~\cite{SI}-- future analysis will increase the agreement of the other variables measured in experiment, and reveal some physics left out in the simplest nontrivial approximations that are made in this paper.

Polarimetry of HSG spectra has given us several unique insights into quantum phenomena in driven solids: in previous works HSG polarimetry has been shown to be a probe of Berry curvatures \cite{DBRprx} and lightwave vallytronics \cite{HSGtmdc}, and has been used to reconstruct Bloch wavefunctions \cite{BlochWave}. HSG is a complement to HHG in solids, which has provided insight into the ultrafast behavior of charges in quantum condensed matter systems ~\cite{AttoSShhg,UltrafastMetrology,QPathHHG}. Recall that in HHG one laser is used to drive both intraband and interband currents and polarizations \cite{HHGintraBand}. The use of two frequencies in HSG enables one to create carefully selected superpositions of charged quasiparticles on demand by controlling the polarization and frequency of the NIR laser, and to independently choose the field and frequency of the strong THz laser that accelerates these quasiparticles \cite{Qile}. By leveraging this high degree of control, we have been able to create, as reported in this Letter, a Michelson interferometer for Bloch waves. The analytical model directly connects the experimentally observed sideband polarizations with material parameters and laser fields and frequencies. Future experiments could leverage this connection to reconstruct materials parameters including the gap through the $\Delta_{NIR}$ term, the parameters of the effective Hamiltonian for holes which determine $\mu_{\nu}$ and $A_{\nu}$ for the excited quasiparticles, and mechanisms for dephasing via phonons or other scattering mechanisms. Bloch wave interferometry has the potential to become an important new tool for determining the electronic structure of strongly driven quantum condensed matter.

If we look beyond HHG and HSG, the study of driven quantum matter is an extremely active and exciting field, which currently spans topics including  transient light-induced phases of strongly correlated electronic matter \cite{Fausti,Zongarticle}, dynamical localization \cite{Cao2022NatPh..18.1302C}, and time crystals \cite{KhemaniPhysRevLett.116.250401,Time-Crystal-GoogleQAI}. In the context of this broad field, the results of this Letter contribute in two ways. First, the HSG experiments discussed here represent dynamics of an open quantum system occurring over about 20,000 cycles of the strong driving field, and thus probe a highly non-trivial asymptotic quantum state rather than a transient state of a strongly-driven system. Second, this state is very far from thermal equilibrium and yet can be well-understood using a simple analytical model.

\section{Acknowledgements}
The work reported here was funded by NSF DMR 2004995. We thank David Weld for helpful discussions, and Audrina Sewal for participation in experiments.

\bibliography{apssamp}

\end{document}


\preprint{APS/123-QED}

\title{Bloch-Wave Interferometry of Driven Quasiparticles in Bulk GaAs:  \\ Supplementary Information}

\author{Seamus D. O'Hara}

\affiliation{Physics Department, University of California, Santa Barbara, California 93106, USA}
\affiliation{Institute for Terahertz Science and Technology, University of California, Santa Barbara, California 93106, USA}
\author{Joseph B. Costello}
\affiliation{Physics Department, University of California, Santa Barbara, California 93106, USA}
\affiliation{Institute for Terahertz Science and Technology, University of California, Santa Barbara, California 93106, USA}
\author{Qile Wu}
\affiliation{Physics Department, University of California, Santa Barbara, California 93106, USA}
\affiliation{Institute for Terahertz Science and Technology, University of California, Santa Barbara, California 93106, USA}

\author{Ken West}
\affiliation{Electrical Engineering Department, Princeton University, Princeton, New Jersey 08544, USA}

\author{Loren Pfeiffer}
\affiliation{Electrical Engineering Department, Princeton University, Princeton, New Jersey 08544, USA}

\author{Mark S. Sherwin}

\affiliation{Physics Department, University of California, Santa Barbara, California 93106, USA}
\affiliation{Institute for Terahertz Science and Technology, University of California, Santa Barbara, California 93106, USA}

\date{\today}

\maketitle
\section{Sample Preparation and Characterization}

Sidebands were generated from a 500 nm thick bulk gallium arsenide (GaAs) epilayer grown via molecular beam epitaxy. The GaAs epilayer came from the same GaAs wafer used in Ref.~\cite{BlochWave} and the sample was prepared in the same way as described in Ref.~\cite{BlochWave}. One emphasis of the sample fabrication was to make sure that the GaAs epilayer was uniformly and minimally strained. Another goal was to make the sample transmissive to sidebands in the near-infrared (NIR) to visible spectral range and reflective to the terahertz (THz) field to enhance the THz field strength at the GaAs epilayer.

\subsection{Optical Response of Sample}

Following the transfer method described in Ref.~\cite{EpiLiftoff,EpiBond,Cole2013,Cole2016}, the GaAs epilayer was transferred to a sapphire substrate, which was chosen because its thermal expansion coefficient is close to that of GaAs. Before high-order sideband generation (HSG) experiments were run, the sample was mounted in the cryostat, attached to the cold finger with a thermally conductive and vacuum-safe adhesive, and cooled down to 35 K. Cooling down to cryogenic temperatures can induce some strain in the GaAs epilayer, which can distort the band structure of GaAs and introduce non-Abelian Berry curvatures into the dynamics of electron-hole pairs in the HSG experiment~\cite{BlochWave}. Minimizing the strain in the sample is important to minimize Berry Curvatures, where are not included in the Bloch-wave interferometry picture presented here. Making the strain uniform over the size of the NIR beam spot is important to ensure that sideband polarization does not vary across the beam spot.

\begin{figure*}
    \centering
    \includegraphics[scale = 0.95]{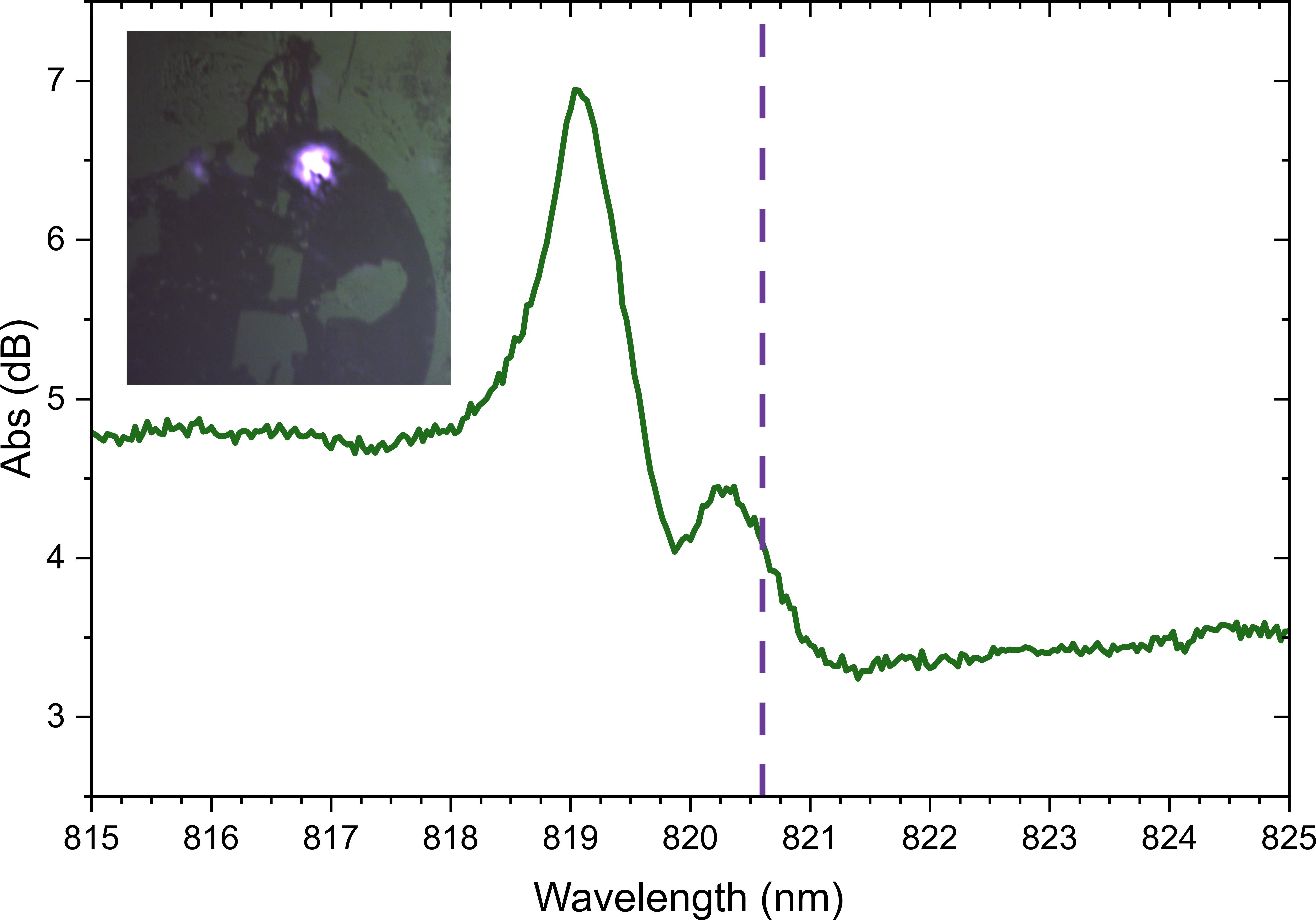} 
    \caption{An absorbance spectrum of the GaAs epilayer. The purple dashed line indicates the NIR wavelength used in experiments reported in this experiment. The measurement was taken at the spot illuminated by a white light source (inset).}
    \label{fig:Abs}
\end{figure*}

To ensure that the NIR beam spot covers an area with a sufficiently uniform and minimal strain, absorbance measurements were performed with a white light source to examine the local strain environment of the GaAs epilayer under the conditions for the HSG experiment. Figure~\ref{fig:Abs} shows an example of an absorbance spectrum, where there are two absorbance peaks associated with different hole spins. The peak splitting arises from the lifting of degeneracy between the heavy-hole (HH) and light-hole (LH) bands due to strain. Such strain-induced exciton splittings were observed in all low-temperature measurements of the absorbance of bulk GaAs films glued or bonded to substrates as far as we know from the literature---see, for example, Ref.~\cite{Glutsch_PhysRevB.50.17009}. After one absorbance spectrum was recorded, the sample was translated by a fraction of the size of the white light beam spot and the absorbance was measured again. This process is repeated over a 7 mm$\times$3 mm section of the sample to find a region with the highest absorbance peak and the most uniform absorbance peak splitting its nearby regions. The region chosen for the HSG experiment is displayed in the inset of Fig.~\ref{fig:Abs}, along with the absorbance spectrum.

As a result of the small strain-induced splitting between LH and HH bands ($\delta_{\rm{HH-LH}} < 3$ meV) shown in Fig. \ref{fig:Abs}, the detunings $\Delta_{\rm NIR}$ are different for electron-hole (E-H) pair excitations associated with different hole spins. The magnitude of such differences are on the same scale as $\delta_{\rm{HH-LH}}$, much less than typical sideband offset energies. For this reason, in this paper, we simply set $\Delta_{\rm NIR}=0$. In a future paper, we will attempt to separately extract the gaps for the E-HH and E-LH pairs.

\subsection{THz field at GaAs epilayer}

In order to enhance the THz field strength at the GaAs epilayer, an indium tin oxide (ITO) layer was deposited onto a clean 488 $\mu$m thick c-axis grown sapphire wafer byusing electron-beam deposition, as detailed in Ref.~\cite{BlochWave}.

Because of the importance of the THz field strengths to our results, care was taken to track the THz field strength during the HSG experiment. The THz field strength was constantly monitored by a pyroelectric detector, whose response was calibrated by a Thomas-Keating (TK) absolute power/energy meter before HSG experimentation each day. A beam splitter directed 10$\%$ of the THz field into the pyroelectric detector, allowing for pulse-by-pulse monitoring while HSG experiments were performed. After the THz beam passed the position at which the TK meter was placed, the THz field was reflected by a flat mirror, a gold-coated off-axis parabolic mirror, and an ITO slide, then transmitted though the cryostat window, and reflected by the ITO coating on the sapphire substrate to build up field enhancement at the GaAs epilayer. In the calculation of the THz field strength, we assume the gold-coated off-axis parabolic mirror and flat mirror are $100\%$ reflective, the ITO slide $70\%$ reflective, the cryostat window $95\%$ transmissive, and the ITO coating on the sapphire provides a $150\%$ field enhacement at the GaAS epilayre. This gives a relative field strength of 99.75$\%$ at the GaAs epilayer with respect to that deduced from the pyroelectric detector signal (see Ref.~\cite{BlochWave} for details).

Two wire gird polarizers were used to attenuate the THz field at the GaAs epilayer to $75 \%$ and $50 \%$ of the peak field strength $F_{\rm THz,0}=70$ kV/cm by adjusting the relative angle, $\Phi$, between the two wire grids, while maintaining the horizontal polarization of the THz field. The equation $F_{\rm THz}(\Phi) = F_{\rm THz,0}\cos^2(\Phi)$ was used to calculate the THz field strength at the GaAs epilayer for an angle $\Phi$ set by a motorized rotation mount. Prior to the HSG experiment, the TK power detector, placed downstream of the wire grid polarizers, was used to calibrate the motor settings and determine the configuration for $\Phi = 0$.

\section{Stokes Polarimetry}

The polarization of the sideband electric fields were measured by using Stokes Polarimetry. We obtained, for each sideband, four Stokes parameters, from which the sideband electric field was represented in a basis of circularly polarized fields, as shown in Fig. 1 (d) of the main text.

\subsection{Optical Setup}

The optical setup is detailed in Ref.~\cite{DBRprx,BlochWave}. The NIR and THz lasers were simultaneously and collinearly focused onto the GaAs sample, leading to generation of high-order sidebands. The sidebands were transmitted out of the cryostat and propagated through the Stokes polarimeter, which consists of a quarter-wave plate (QWP) mounted on a rotation stage and a horizontal linear polarizer.

The sidebands are then coupled into a grating spectrometer and detected with an electron-multiplied CCD (charge-coupled device) camera. The lowest orders of sidebands are cut off by a short-pass optical filter to filter out the NIR laser, which is about $10^{5}$ more intense than the strongest sideband fields. For the experiments reported in this Letter, the 10th-order sideband is the lowest-order sideband to be resolved, and the 36th-order is the highest.

The relative intensities of the sidebands measured through the CCD camera, $S_{0,{\rm out}}$, are dependent on the Stokes parameters, $S_{i}$ ($i=0,1,2,3$), and the orientation angle of the QWP fast axis with respect to the horizontal,  $\theta_{\rm QWP}$, following the expression
\begin{align} \label{Sout}
    S_{0,out}(\theta_{\rm QWP}) 
    = & \frac{S_{0}}{2} + \frac{S_1}{4} - \frac{S_3}{2}\sin{2\theta_{\rm QWP}}\notag\\
     &+ \frac{S_1}{4}\cos{4\theta_{\rm QWP}} + \frac{S_2}{4}\sin{4\theta_{\rm QWP}}.
\end{align}
For a given Stokes polarimetry experimental run, $S_{0,\mathrm{out}}$ of each sideband was measured with the angle $\theta_{\rm QWP}$ sweeping from 0 to 360 degrees in 22.5-degree steps. The four Stokes parameters were then calculated from Fourier transforms of $S_{0,{\rm out}}$ with respect to the angle $\theta_{\rm QWP}$, as detailed in Ref.~\cite{BlochWave}.

\subsection{Representation of the Sideband Electric Field}

To conveniently make a comparison between theory and experiment, we choose the basis of circular polarization fields as
\begin{align}
       |R\rangle=\frac{\hat{X}+i\hat{Y}}{\sqrt{2}},\\
       |L\rangle=-\frac{\hat{X}-i\hat{Y}}{\sqrt{2}},
\end{align}
with $\hat{X}$ ($\hat{Y}$) being a unit vector along [001] ([010]) crystal axis of bulk GaAs, and write the polarization state of the $n$th-order sideband as
\begin{align}
|E_{\rm SB,n} \rangle= l(n)e^{i\phi(n)}|L\rangle+ r(n)|R\rangle,
\end{align}
where the functions $l(n)$, $r(n)$, and $\phi(n)$ are all real, and satisfy $l^2(n)+r^2(n)=1$.

\begin{figure}
    \centering
    \includegraphics[width=0.47\textwidth]{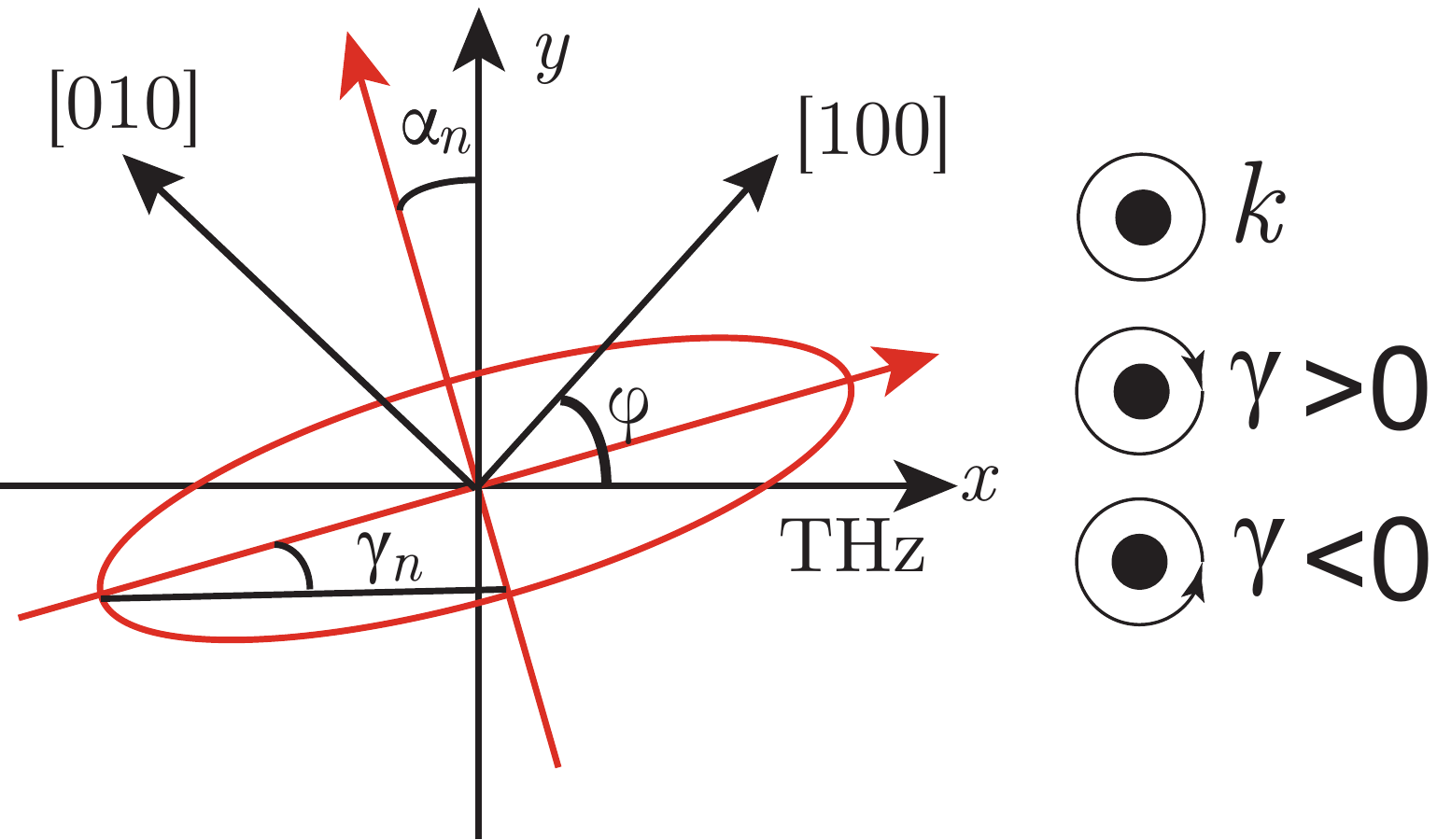} 
    \caption{Definition of the orientation angle $\alpha_n$ and ellipticity angle $\gamma_n$. The sign of the ellipticity angle $\gamma_n$ is defined with respect to the wave vector $k$.}
    \label{fig:alpha_gamma}
\end{figure}

The polarization state $|E_{\rm SB,n} \rangle$ can be determined by the Stokes parameters through the orientation angle $\alpha_n$ and ellipticity angle $\gamma_n$ defined with respect to the horizontal polarization of the THz field (see Fig.~\ref{fig:alpha_gamma} for the definition). The Jones vector associated with the polarization state $|E_{\rm SB,n} \rangle$ can be written as
\begin{align}
      \begin{pmatrix}
      E_x \\
      E_y
      \end{pmatrix}
      \propto
      \begin{pmatrix}
      \cos\alpha\cos\gamma-i\sin\alpha\sin\gamma\\
      \sin\alpha\cos\gamma+i\cos\alpha\sin\gamma
      \end{pmatrix},
\end{align}
with the $x$ axis defined by the horizontal polarization of the THz field. The Jones vector in the basis $\{|R\rangle, |L\rangle\}$ can be related to the Jones vector $(E_x,E_y)^{T}$ through a unitary transformation:
\begin{align}
      \begin{pmatrix}
      r(n) \\
      l(n)e^{i\phi(n)}
      \end{pmatrix}
      \propto
      \frac{1}{\sqrt{2}}
\begin{pmatrix}
e^{i\varphi} & -ie^{i\varphi}\\
-e^{-i\varphi} & -ie^{-i\varphi}
\end{pmatrix}
      \begin{pmatrix}
      E_x \\
      E_y
      \end{pmatrix}
\end{align}
where $\varphi$ is the angle between the [100] crystal axis and the linear polarization of the THz field. After some algebra, we obtain
\begin{align}
    r^2(n) &= \frac{1+\sin2\gamma}{2}, \\
    l^2(n) &= \frac{1-\sin2\gamma}{2}, \\
    \phi(n) &=2(\alpha-\varphi) +  \pi\, ({\rm mod}\,2\pi).
\end{align}

Using the relations connecting the Stokes parameters with the angles $\alpha$ and $\gamma$~\cite{BlochWave},
\begin{align}
    \sin2\gamma &= \frac{S_3 }{S_1^2+S_2^2+S_3^2}, \\
    \tan2\alpha &=\frac{S_2}{S_1},
\end{align}
we arrive at the equations
\begin{align}
    r^2(n) &= \frac{1+\tilde{S}_3}{2}, \\
    l^2(n) &= \frac{1 - \tilde{S}_3}{2}, \\
    \phi(n) &= \arctan\frac{\tilde{S}_2}{\tilde{S}_1} - 2\theta -  \pi/2\, ({\rm mod}\,2\pi),
\end{align}
where $\theta$ is the angle between the [110] crystal axis and the linear polarization of the THz field, and the Stokes parameters, $S_1$, $S_2$, and $S_3$, are normalized as $\tilde{S}_j=S_j/\sqrt{S_1^2 + S_2^2 + S_3^2}$ ($j=1,2,3$). For the HSG experiments reported in this work, the angle $\theta$ is measured to be 88 degrees.

\section{Theory of Sideband Polarization}

To calculate the sideband polarization, we need a model for the propagators describing the trajectories taken by the E-H pairs. Following Ref. \cite{BlochWave}, we start with the expression for the propagators.

\begin{align}\label{Quantum}
    \varsigma_n^{{\rm HH}({\rm LH})} &= \int_{0}^{\frac{2\pi}{\omega}} dt e^{i(\Omega+n\omega)t} \cdot \notag\\
    &\int_{-\infty}^{t}dt'e^{i[-E_{\rm g}(t-t')/\hbar+A_{{\rm HH}({\rm LH})}(t,t')]-(\Gamma_d/\hbar)(t-t')}e^{-i\Omega t'},
\end{align}
where $\Omega$ and $\omega$ are the angular frequencies of the NIR laser and THz field, respectively, $\Gamma_d$ is the dephasing constant, $\hbar$ is the reduced Planck constant, and 
\begin{align}\label{ALIT}
A_{\rm{HH(LH)}}(t,t') =  -\int_{t}^{t'}dt''\frac{\hbar{{\bf k}^2(t'')}}{2\mu_{\rm HH(LH)}}
\end{align}
is the dynamic phase acquired by the E-H pairs from the time $t$ to $t'$ with the contribution from the bandgap $E_{\rm g}$ subtracted, and $\mu_{\rm HH}$ ($\mu_{\rm LH}$) is the reduced mass of the E-HH (E-LH) pairs. The integral includes all recollision pathways starting from ${\bf k}=0$. 

\subsection{Semiclassical Theory of SB Polarization}
The expression for $\varsigma_n^{\nu}$ in Eq. \ref{Quantum} is a full quantum path integral which allows for any creation time $t$ and recombination time $t'$ to contribute to the overall sideband polarization. Given the moderately strong dephasing rates between quasiparticles and the GaAs lattice, we assume the leading contributions to are from the quasiparticles accelerated for less than one THz cycle.  Taking the integration variables $t'$ and $t$ as the creation and recombination times associated with a
shortest recollision pathway, $t_{o,n,\nu}$ and $t_{f,n,\nu}$ ($\nu={\rm HH,LH}$), and neglecting all quantum fluctuations, we model the propagators as
\begin{align}\label{Semi2}
     \varsigma_{n}^{\nu} 
     = e^{i[
     n\omega t_{f,n,\nu} 
     + A_{n,\nu}
     +(i\Gamma_d +\Delta_{\rm NIR}) 
     \tau_{n,\nu}/\hbar
     ]},
\end{align}
where $\tau_{n,\nu} \equiv t_{f,n,\nu} - t_{o,n,\nu}$ is the time duration, $\Delta_{\rm NIR}=\hbar\Omega-E_{\rm g}$ is the detuning of the NIR laser, and $A_{n,\nu}=A_{\nu}(t_{f,n,\nu},t_{o,n,\nu})$. Here, the recollision processes are considered to be fully classical. We set $\Delta_{\rm NIR}=0$ for simplicity as mentioned earlier.

\subsection{Classical Conundrum}

To obtain the creation and recombination times, $t_{o,n,\nu}$ and $t_{f,n,\nu}$ ($\nu={\rm HH,LH}$), we start with the Newton's equation of motion, $\hbar\dot{k}=-eE_{\rm THz}(t)$, with $e$ being the elementary charge and  a sinusoidal THz field in the form $E_{\rm THz}(t) = F_{\rm THz}\sin(\omega t)$. Taking the initial condition $k(t_{o,n,\nu}) = 0$, we obtain the solution for the momentum,

\begin{align}\label{kSin}
        \hbar k(t) = \frac{eF_{\rm THz}}{\omega}
        [\cos(\omega t)-\cos(\omega t_{o,n,\nu})].
\end{align}
The positions of the electrons and holes are then given by
\begin{align}\label{xSin}
        x(t) = &\int_{t_{o,n,\nu}}^t dt''\frac{\hbar k(t'')}{m}\notag\\
        =&\frac{-eF_{\rm THz}}{m\omega^2} [\omega(t-t_{o,n,\nu})\cos(\omega t_{o,n,\nu}) \notag\\
        &+\sin(\omega t_{o,n,\nu}) - \sin(\omega t)],
\end{align}
where $m$ is the effective mass. Using the energy conservation equation, $\hbar^2k^2(t_{f,n,\nu})/(2\mu_{\nu}) = n\hbar\omega$, and the condition that, at recollision, the electrons and holes return to the origin, $x(t_{f,n,\nu}) = 0$, one can solve the times $t_{o,n,\nu}$ and $t_{f,n,\nu}$ numerically (see, for example, Ref.~\cite{Qile}).

\subsection{The linear-in-time (LIT) approximation}

\begin{figure}
    \centering
    \includegraphics[scale = 0.88]{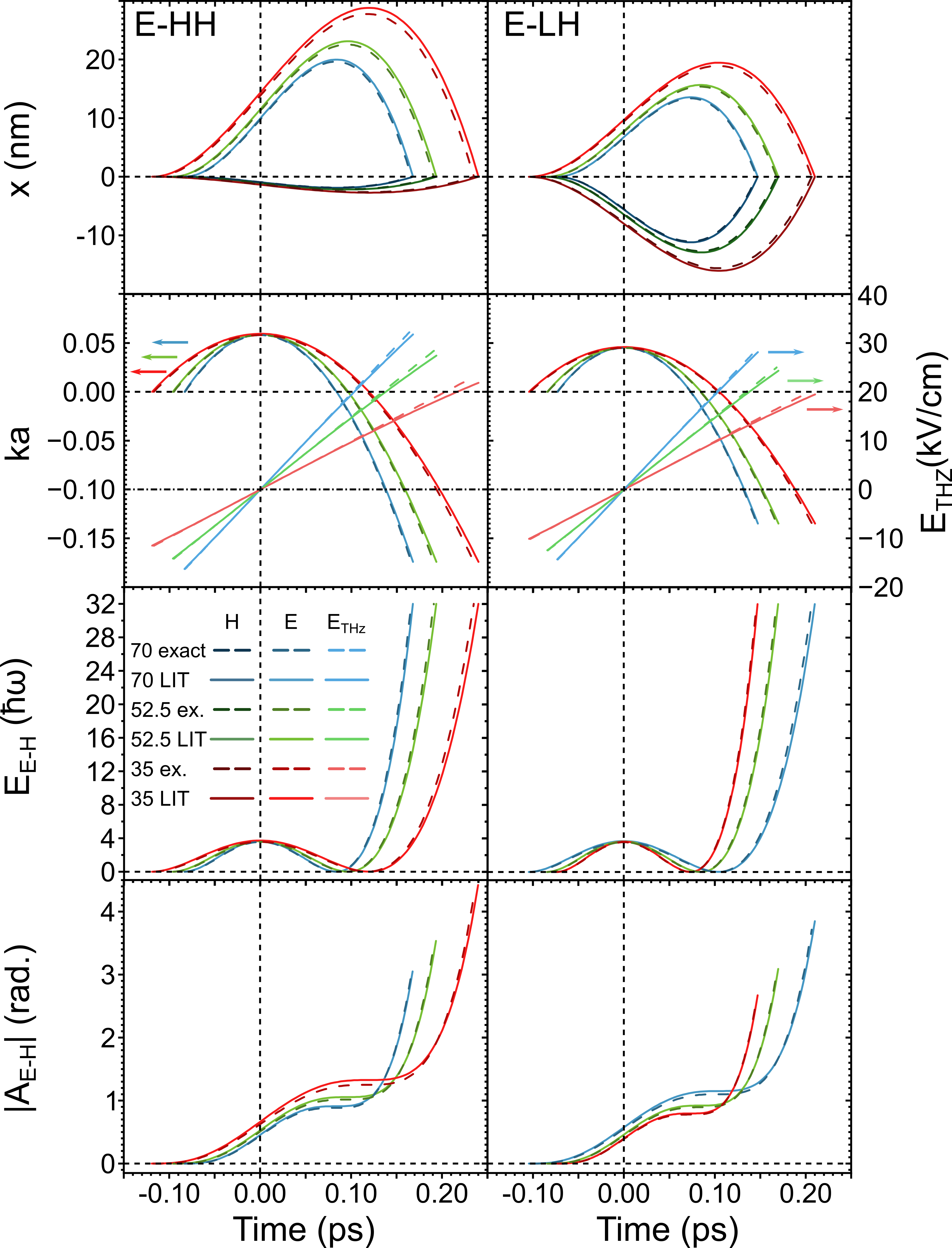}
    \caption{Values calculated from LIT (dashed line) and numerical solutions of a sinusoidal (solid line) THz field for E-H pairs which produce a 32nd order sideband. The different field strengths used in experiment, 70 kV/cm (blue), 52.5 kV/cm (green), and 35 kV/cm (red), are all displayed. The left column holds the values for the E-HH pairs, and the right holds the values for the E-LH pairs. The top row holds the calculated x-position for the holes (dark colors) and electrons (light colors). The second row hold the calculated wave vectors for the electrons (downward parabolas, associated with the left y-axis, a = 5.56 $\AA$) and the THz fields (lines, associated with the right y-axis). The third row holds the calculated E-H pair kinetic energy. The bottom row holds the calculated absolute values of the E-H pair dynamical phase. }
    \label{fig:HHcompare}
\end{figure}

To further simplify the analysis, we apply a Taylor expansion approach to the THz field. Given that all observed sidebands can be associated with electron-hole pairs that recollide less than 250 fs after a node in the THz field (see Fig \ref{fig:HHcompare}),  whereas the THz field used in experiment has a period of 2.22 ps, we make the approximation that the THz field is linear in time (LIT):

\begin{align} \label{LIT}
    E_{\rm THz}(t) \approx F_{\rm THz} \omega t.
\end{align}
The LIT approximation has been applied to solve the semiclassical trajectories of E-H pairs in HSG~\cite{Qile}. 
For the special case of classical recollisions, the creation and recombination times, $t_{o,n,\nu}$ and $t_{f,n,\nu}$ ($\nu={\rm HH,LH}$) satisfy~\cite{Qile}
\begin{align} \label{timesLITsimp}
    -2\omega t_{o,n,\nu} = \omega t_{f,n,\nu} = \frac{2}{\sqrt{3}}\left(\frac{2nE_{\rm ph}}{U_{\nu}}\right)^{1/4},
\end{align}
where $E_{\rm ph}$ is the THz photon energy, and $U_{\nu}=e^2F^2_{\rm THz}/(4\mu_{\nu}\omega^2)$ is the ponderomotive energy. Within the LIT approximation, the momentum $\hbar k(t)$ and the positions of the electron-hole pairs are all polynomials as functions of time:
\begin{align} \label{kLIT}
        \hbar k(t) = \frac{eF_{\rm THz}}{2\omega} \left[(\omega t_{o,n,\nu})^2-(\omega t)^2\right],
\end{align}
\begin{align} \label{xLIT}
        x(t)= \frac{-eF_{\rm THz}}{2m\omega^2}\left[(\frac{1}{3}\omega t)^3 - (\omega t_{o,n,\nu})^2\omega t + \frac{2}{3}(\omega t_{o,n,\nu})^3\right].
\end{align}
The solution of $t_{o,n,\nu}$ and $t_{f,n,\nu}$ (Eq. \ref{timesLITsimp}) can be derived from the conditions $x(t_{f,n,\nu}) = 0$ and $\hbar^2k^2(t_{f,n,\nu})/(2\mu_{\nu}) = n\hbar\omega$.

\subsection{Calculation of the Propagators}

With the linear-in-time approximation, using the solutions of the creation and recombination times, $t_{o,n,\nu}$ and $t_{f,n,\nu}$  ($\nu={\rm HH,LH}$), we obtain the dynamic phase using Eq.~\ref{ALIT}:
\begin{align}
    A_{n,\nu} = -\frac{2\sqrt{3}}{15}\left(\frac{8n^5\hbar\omega^3\mu_{\nu}}{e^2F_{\rm THz}^2}\right)^{1/4}.
\end{align}
The propagators are then calculated by using Eq.~\ref{Semi2} as
\begin{align}\label{vsigmaLIT}
        \varsigma_{n}^{\nu} =\exp\left[i\left(\frac{8n}{15}  + \left(\frac{i\Gamma_d +\Delta_{\mathrm{NIR}}}{\hbar\omega}\right)\right)\left(\frac{72n\hbar\omega^3\mu_\nu}{e^2F_{\mathrm{THz}}^2}\right)^{1/4}\right].
\end{align}

\subsection{Sideband Polarization}

Following Ref.~\cite{BlochWave}, we connect the sideband field ${\bf E}_{\rm SB,n}$ with the NIR laser field ${\bf E}_{\rm NIR}$ using the following relation,
\begin{align}
    {\bf E}_{\rm SB,n}
    \propto
    	\sum_{s=1,2}
	&
        		\begin{pmatrix}
	         {\bf d}_{\rm E-HH,s}\\
        		 {\bf d}_{\rm E-LH,s}
	        \end{pmatrix}^{\dag}
	        		\begin{pmatrix}
	         		\varsigma^{\rm HH}_n  & 0\\
	         		0  &  \varsigma^{\rm LH}_n       		 	
	       		 \end{pmatrix} 	      
	  \notag\\          		
	  &
	        \begin{pmatrix}
	         {\bf d}_{\rm E-HH,s}\\
        		 {\bf d}_{\rm E-LH,s}
	        \end{pmatrix} 
	\cdot
	{\bf E}_{\rm NIR},
 	\label{EQ:ESB_from_ENIR}
\end{align}
where $s$ labels the two-fold degeneracy in the electron-hole states, ${\bf d}_{\rm E-HH,s}$ (${\bf d}_{\rm E-HH,s}$) is the dipole vector associated with the E-HH (E-LH) states labeled by $s$. In the basis of circular polarization fields, $\{|R\rangle, |L\rangle\}$, this relation can be written as Jones vectors connected by a dynamical Jones matrix in the form,
\begin{equation} \label{Jones}
    \begin{pmatrix}
        r(n) \\ l(n)e^{i\phi(n)}
    \end{pmatrix} 
    \propto
    \begin{pmatrix}
        T_{++,n} & T_{+-,n} \\
        T_{-+,n} & T_{--,n}
    \end{pmatrix} 
    \begin{pmatrix}
        r_{\mathrm{NIR}} \\ l_{\rm NIR}e^{i\phi_{\rm NIR}}
    \end{pmatrix},
\end{equation}
where $(r_{\rm NIR}, l_{\rm NIR}e^{i\phi_{\rm NIR}})^T$ is the Jones vector for the NIR laser. The dynamical Jones matrix elements are calculated as~\cite{BlochWave}
\begin{align} \label{Tsigma}
        T_{++,n} = T_{--,n} &= \frac{2 +n _z}{3} \varsigma_{n}^{{\rm HH}} +  \frac{2 -n_z}{3} \varsigma_{n}^{{\rm LH}}, \\
        T_{+-,n} &= \frac{n_x + in_y}{\sqrt{3}}\left(\varsigma_{n}^{{\rm HH}} - \varsigma_{n}^{{\rm LH}}\right), \\
        T_{+-,n} &= \frac{n_x - in_y}{\sqrt{3}}\left(\varsigma_{n}^{{\rm HH}} - \varsigma_{n}^{{\rm LH}}\right).
\end{align}
where $(n_x,n_y,n_z)$ is a unit vector along the vector $((\sqrt{3}/2)\sin2\theta,-({\sqrt{3}\gamma_3}/{2\gamma_2})\cos2\theta,-{1}/{2})$. Since we have an analytical form of the propagators $\varsigma_n^{\nu}$ ($\nu={\rm HH,LH}$) from Eq. \ref{vsigmaLIT}, the dynamical Jones matrix elements can also be calculated analytically. The dashed lines presented in Figure 2 in the main text are produced by using these analytical expressions.

\subsection{Accuracy of LIT}

We compare the values of position $x(t)$, wave vector $k(t)$, kinetic energy $E_{e-\nu}(t)=\hbar^2 k^2(t)/(2\mu_{\nu})$, and dynamic phase $A_{n,\nu}$ ($\nu={\rm HH, LH}$), calculated by using the LIT approximation and the numerical solutions by using $E_{\mathrm{THZ}}(t) = F_{\mathrm{THz}}\sin(\omega t)$ (sinusoidal THz field). Figure~\ref{fig:HHcompare} demonstrates the accuracy of the LIT approximation for time domains relevant to the sideband generation observed in the experiment.

\bibliography{apssamp}